\documentclass[10pt,conference]{./sty/IEEEtran}

\usepackage{subfigure}
\usepackage{color}
\usepackage{dsfont}
\usepackage{cite}
\usepackage{graphicx}
\usepackage{subfigure}
\usepackage{url}
\usepackage{amsmath}
\usepackage{balance}

\begin{document}

\title{When Network Coding and Dirty Paper Coding meet in a Cooperative Ad Hoc Network}

\author{
\authorblockN{Nadia Fawaz, David Gesbert}
\authorblockA{Mobile Communications Department, Eurecom Institute \\
Sophia-Antipolis, France \\
\{fawaz, gesbert\}@eurecom.fr}
\and
\authorblockN{Merouane Debbah}
\authorblockA{Alcatel-Lucent Chair on Flexible Radio, Sup\'{e}lec\\
   Gif-sur-Yvette, France\\
   merouane.debbah@supelec.fr}
}

% make the title area
\maketitle

\begin{abstract}
We develop and analyze new cooperative strategies for ad hoc networks that are more spectrally efficient than classical DF cooperative protocols.
Using analog network coding, our strategies preserve the practical
half-duplex assumption but relax the orthogonality constraint. The introduction of interference due to non-orthogonality is mitigated thanks to precoding, in particular Dirty Paper coding. Combined with smart power allocation, our cooperation strategies allow to save time and lead to more efficient use of bandwidth and to improved network throughput with respect to classical RDF/PDF.
\end{abstract}

\section{Introduction}\label{sec:Introduction}

\begin{figure}[htbp]%[htbp]
  \begin{center}
  \includegraphics[scale=1]{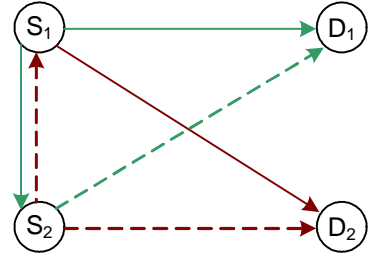}\\
  \caption{A four node network with 2 cooperating sources and 2 destinations}
  \label{fig:fourNodeNetwork}
  \end{center}
\end{figure}

\PARstart{C}{ooperative} communications occur when distributed
wireless nodes interact to jointly transmit information. Several radio terminals relaying signals for each other form a
virtual antenna array
and their cooperation enables
the exploitation of spatial diversity in fading channels.  Several
relaying strategies already exist, the simplest
and most famous being \cite{Laneman-2004} Amplify and Forward (AF) and Decode and Forward (DF) with repetition coding (RDF) or parallel channel coding (PDF).
Since radio terminals cannot transmit and receive simultaneously in
the same frequency band, most cooperative strategies are based on
half-duplex mode. When considering a three-node cooperative network,
with a source S, a relay R and a destination D, each transmission is
divided into two blocks: in first block, S transmits and R and D receive; in second block R relays and D receives. In some strategies S transmits also in second block.
Now let us consider the four-node network in fig.
(\ref{fig:fourNodeNetwork}) with two sources $S_1$ and $S_2$
transmitting in a cooperative fashion to two destinations $D_1$ and
$D_2$ as in \cite{Laneman-2004}. The previous transmission scheme is
repeated twice, first for the relay channel $S_1 - S_2 - D_1$ and
second for the relay channel $S_2 - S_1 - D_2$ as described in fig.
\ref{fig:SimultaneousTxRelay} (b), resulting in four-block
transmission. The use of orthogonal interference free channels for
sources and relays transmissions simplifies receiver algorithms but
results in a loss of bandwidth.

\subsection{The Idea in Brief}

\begin{figure*}[htbp]%[htbp]
  \begin{center}
  \includegraphics[scale=1]{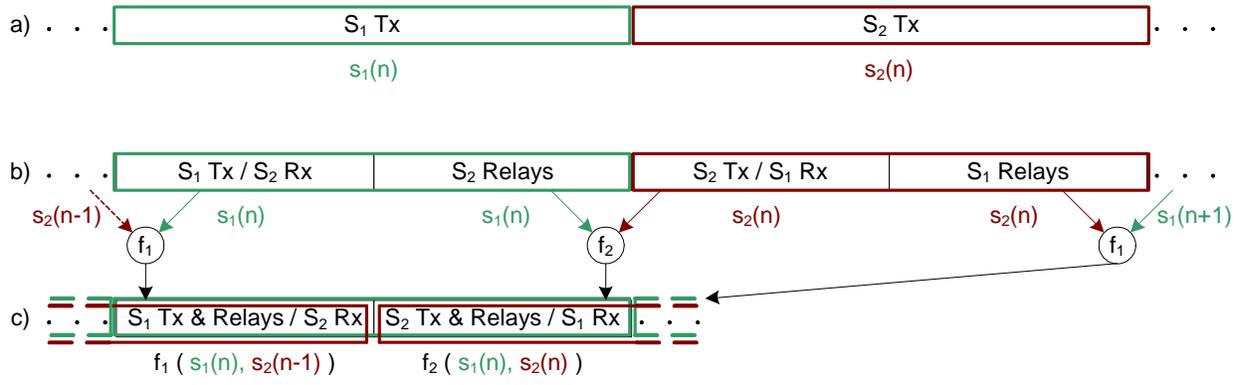}\\
  \caption{Time division channel allocations for (a) orthogonal direct transmissions,
  (b) usual orthogonal cooperative transmissions (c) proposed scheme : analog network coding cooperative transmissions}
  \label{fig:SimultaneousTxRelay}
  \end{center}
\end{figure*}

Loss of bandwidth issue has been tackled at higher layers thanks to
network coding (NC). Packets arriving at a node on any edge of a network
are put into a single buffer. At each transmission opportunity, an
output packet is generated as a random linear combination of packets
in the buffer within "current" generation \cite{Chou-tutorial}.
Inspired by network coding, consider
a four-node cooperative network
using "network precoding" in a two-block transmission scheme, where
in each single block one source simultaneously transmits and relays
as in fig. \ref{fig:SimultaneousTxRelay} (c):
\begin{itemize}
  \item first block : $S_1$ sends a single signal $f_1(s_1(n),s_2(n-1))$ which is a function of
  both its own message $s_1(n)$ and a message $s_2(n-1)$ received, decoded and re-encoded by $S_1$ in the
  second block of previous transmission (repetition of the codeword - RDF -  or use of an independent codeword -PDF), now relayed for $S_2$. $S_2$, $D_1$ and $D_2$ receive.
  Since $S_2$ knows the message in $s_2(n-1)$, it can extract $s_1(n)$, if it also knows the mixing function $f_1$.
  \item second block : $S_2$ sends a single signal $f_2(s_2(n),s_1(n))$ which is a function of
  both its own message $s_2(n)$ and a message $s_1(n)$ received, decoded and re-encoded by $S_2$ in the
  first block of the current transmission, now relayed for $S_1$. $S_1$, $D_1$ and $D_2$
  receive. Since $S_1$ knows the message in $s_1(n)$, it can extract $s_2(n)$, if it also knows $f_2$.
\end{itemize}
Functions $f_1$ and  $f_2$ are the network precoding functions which
help improving communication in terms of bandwidth. Knowing $f_1$
and $f_2$ allows sources $S_2$ and $S_1$ to easily cancel
interference and extract the message they will have to relay in next
block. But unfortunately, bandwidth usage improvements have a cost: the
introduction of interference at destinations $D_1$ and $D_2$ . In
first block, $s_2(n-1)$ is intended to $D_2$ as relayed signal and
acts as interference for $D_1$, which is only interested in
$s_1(n)$; reciprocally, $s_1(n)$, intended to $D_1$, generates
interference for $D_2$ interested in $s_2(n-1)$. A similar
interference problem occurs in second block. Nevertheless,
interference is known at transmitter, thus one can design the
precoding functions to take into account this issue. In particular
Dirty Paper Coding (DPC) \cite{Costa-1983}, a well-known coding technique to mitigate interference known at transmitter,  may help NC.
We may expect DPC-like network precoding to help improving bandwidth
efficiency in a cooperative network as well as mitigating
interference, thus enhancing performance with respect to usual cooperative schemes.

\subsection{Related Work}

In \cite{Sendonaris-Erkip-2003} a cooperation strategy is proposed
for two transmitters and one destination. Each
source transmits both information of its own and of its partner,
orthogonally superposed
using orthogonal spreading codes leading to improved user capacity.
Nevertheless, a common destination is assumed for the
cooperating pair, the half-duplex constraint is not taken into
account,
and cooperative periods are divided into two parts: slots where sources
transmit only their own signal and slots where they send a
cooperative signal. Our proposed scheme is more
efficient, because
no orthogonality constraint is imposed for source and relayed signal separation.
In \cite{Hunter-Sanayei-2006} coded cooperation (CC) is introduced in
a system with two sources and one destination and is shown to
outperform AF and RDF.
In that scheme, frame separation of own and relayed signals again leads to bandwidth loss and a common destination is assumed, a particular case of cooperative
system.
In\cite{Azarian-ElGamal-2005} non-orthogonal AF (NAF)
protocols - yet preserving the half-duplex constraint - are proposed. In NAF,
orthogonality constraint is relaxed by letting the source transmit
symbols even when a relay is retransmitting. NAF turns out to
improve performances with respect to classical AF.
Nevertheless with NAF, only half of the symbols are relayed. In our scheme, orthogonality between source and
relayed signals is also relaxed, half-duplex preserved, but all
symbols benefit from cooperative transmission.
All these works consider a common destination and do not address
interference mitigation issues arising in multi-source
multi-destination cooperative ad hoc system.

DPC was considered in relay networks in \cite{Ng-Goldsmith-2004},
\cite{Lo-2005} and \cite{HostMadsen-2006}.
In \cite{Ng-Goldsmith-2004} DPC transmit cooperation scheme suffers
from loss of bandwidth due to the orthogonal cooperation channel
used to exchange transmit messages between the two sources and whose
cost is not taken into account.
In \cite{Lo-2005}, a full duplex S-R-D network is considered, in
which the source S sends a signal consisting of two components, one
intended to the relay and one intended to the destination. In
this relay network, DPC precoding is used at source to mitigate the
interference caused at the relay by the second component.
On the contrary, in our cooperation scheme, NC takes care of interference at the relay,
whereas DPC is used at source and at relay to mitigate interference caused at destinations.
In \cite{HostMadsen-2006} DPC is considered for full-duplex transmit cooperation, with the sources jointly deciding the codewords both will combine in their transmit signals, which needs some signaling to agree on the codewords, not taken into account in the resource expenses. Besides the DPC-ordering
is fixed before power allocation optimization, which impacts the individual rates and makes one destination use forward-decoding and the other backward-decoding. On the contrary, as in \cite{Laneman-2004} we consider a TDMA scheme, but with a time shift between the decoding of received signals at destinations, allowing to respect the half-duplex constraint, while NC allows to maintain a continuous flow of information interesting both destinations. Therefore our strategies are the first to manage combining the half-duplex constraint in the \cite{Laneman-2004}-fashion and the continuous transmission of data interesting all destinations in the \cite{HostMadsen-2006}-way. Moreover in our scheme, each source chooses its codewords alone, without needing to know what the other chose and both sources select the best DPC-orderings as part of the optimization, which they can achieve alone as long as channel information is available. Finally both destinations can use forward-decoding and do not to need to wait until the end of a frame of codewords to decode backward the first codeword sent.

The idea of analog network coding at the physical layer was proposed in \cite{Fawaz-Gesbert-Debbah-WSIT2007} with power allocation, interference mitigation tanks to DPC and results on the total network throughput, nevertheless the full analysis is  presented in this paper.
Recently \cite{Katti-Maric-Goldsmith-2007} studied AF with analog network coding and showed that joint relaying and network coding can enhance the network throughput.

Our main contribution is to bring network coding, in an analog way, at the physical layer, to provide novel cooperative protocols using analog network coding and to analyze their performances in terms of the network throughput and outage behavior. Thanks to analog Network Coding combined with Dirty Paper precoding, time is saved compared to classical DF protocols, interference resulting from non-orthogonality is mitigated, leading to a better use of ressources and improved spectral efficiency. Analysis show that our cooperative strategies clearly outperform classical orthogonal DF protocols.

\subsection{Outline}

The rest of the paper is organized as follows. In section
\ref{sec:SysMod}, notations and the system model are
presented. In section \ref{sec:PrecodMeth}, cooperative precoding
methods are described whereas the performance criteria are derived
in section \ref{sec:PerfEval}. Numerical results and comparison with
other cooperative protocols are provided in section \ref{sec:NumRes}
and lead to the concluding section \ref{sec:Conclusion}.

\section{System Model}\label{sec:SysMod}

Considering $i\in\{1,2\}$, $\bar{i}$ denotes the
complementary integer in the ensemble, e.g. if $i=1$,
$\bar{i}=2$. Matrices and vectors are represented by boldface
uppercase. $\textbf{A}^T$, $\textbf{A}^\ast$, $\textbf{A}^H$ denote
the transpose, the conjugate and the transpose conjugate of matrix \textbf{A}. tr(\textbf{\textbf{A}}), $\det(\textbf{A})$ and $\|\textbf{A}\|_F =
\sqrt{ tr(\textbf{A} \textbf{A}^H) }$ stand for trace, determinant
and Frobenius norm of \textbf{A}. $\mathds{E}$ is
statistical expectation and $\textbf{R}_\textbf{V} =
\mathds{E}[\textbf{V} \textbf{V}^H]$ is the correlation matrix of
vector $\textbf{V}$. Finally $\textbf{I}_N$ is the identity matrix
of size N.

To capture the gain resulting from the NC
approach, we consider that all terminals are equipped with a
single antenna.
Consider the four node network illustrated in fig.
\ref{fig:fourNodeNetwork}.  Each source $S_i \: , \: i\in\{1,2\}$
generates a sequence $s_i(n)\: , \: n\in\{1,..,N\}$.
These symbols are modeled by independent identically distributed
(i.i.d.) circularly-symmetric complex gaussian random variables,
with zero mean and variance $\varepsilon_s=\mathds{E}[|s_i(n)|^2]$.
At time $t=kT=k/W \:,\:k \in\mathds{N}$, the signal
transmitted by $S_i$ is denoted $x_i(k)$ whereas $y_{S_i}(k)$
and $y_{D_j}(k)$ represent the signals received by source $S_i$ and
destination $D_j$ respectively, with  $i,j\in\{1,2\}$ . Finally
$f_i$ represents the network coding function performed at
$S_i$. Those functions can be of any kind, not necessarily linear.
Nevertheless, in this paper developing a network coding
approach for cooperative ad hoc networks, we focus first on
functions performing a linear operation on the symbols $s_1$ and
$s_2$, to simplify analysis and detection at destinations.
Then a DPC approach is considered and shown to outperform the
other strategies.

As described in section \ref{sec:Introduction} and figure
\ref{fig:SimultaneousTxRelay} (c), NC cooperative
communication divides each transmission into two blocks.
\begin{itemize}
  \item \textbf{First block} at even time indexes $k=2n$, signals transmitted by $S_1$ and received by other terminals are:
  \begin{eqnarray}
    \!\!\!\!\!\!\!\!\! x_1(2n) \!\!\! &=& \!\!\! f_1(s_1(n),s_2(n-1))\nonumber\\
    \!\!\!\!\!\!\!\!\! y_{S_2}(2n) \!\!\! &=& \!\!\! h_{S_2 S_1} \: x_1(2n) + z_{S_2}(2n) \nonumber \\
    \!\!\!\!\!\!\!\!\! y_{D_j}(2n) \!\!\! &=& \!\!\! h_{D_j S_1} \: x_1(2n) + z_{D_j}(2n) \mbox{ , } j\in\{1,2\} \nonumber
  \end{eqnarray}
  \item \textbf{Second block} at odd time indexes $k\!=\!2n\!\!+\!\!1$, signals transmitted by $S_2$ and received by other terminals are:
  \begin{eqnarray}
    \!\!\!\!\!\!\!\!\! x_2(2n+1) \!\!\! &=& \!\!\! f_2(s_1(n),s_2(n)) \nonumber \\
    \!\!\!\!\!\!\!\!\! y_{S_1}(2n+1) \!\!\! &=& \!\!\! h_{S_1 S_2} \, x_2(2n+1) + z_{S_1}(2n+1) \nonumber \\
    \!\!\!\!\!\!\!\!\! y_{D_j}(2n+1) \!\!\! &=& \!\!\! h_{D_j S_2} \, x_2(2n+1) + z_{D_j}(2n+1) \mbox{ , } j\in\{1,2\} \nonumber
  \end{eqnarray}
\end{itemize}

The channel between transmitter $u\in\{S_1,S_2\}$ and receiver
$v\in\{S_1,S_2,D_1,D_2\}$ is represented by $h_{v u}$ which includes
the effects of path-loss, shadowing and slow flat fading. These
channel coefficients are modeled by independent circularly-symmetric
complex gaussian random variables with zero mean and variance
$\sigma_{v u}^2$, i.e. Rayleigh fading. $z_v(k)$ are i.i.d
circularly-symmetric complex gaussian noises at receivers, with
variance $\sigma^2$.
Each source has a power constraint in the continuous time-channel of
P Joules/s and transmits only half of the time, both in orthogonal
interference-free cooperation scheme and in the proposed NC cooperation schemes. Thus the power constraint translates
into $P_i=\mathds{E}[|x_i(n)|^2]\leq \frac{2P}{W}$. Since a source
transmits only part of time, it can increase its transmit power in
its transmission block and remain within its average power
constraint for the whole transmission.

\section{Precoding Method}\label{sec:PrecodMeth}

\subsection{Linear Precoding}

In Linear Network Coding for RDF, $S_1$ detects $s_2(n-1)$ in the
signal transmitted by $S_2$ and re-encodes it using the same
codeword. Then $S_1$ forms its transmitted signal $x_1(n)$ as a
linear combination of its own codeword $s_1(n)$ and the repeated
$s_2(n-1)$. The same process happens at $S_2$.
Therefore function $f_i$ can be represented by a matrix
$\textbf{F}_i$ of size $N_t \times N_s$, i.e. (number of transmit
antennas at source) times (number of symbols on which $f_i$ acts).
In the single antenna scenario, $\textbf{F}_i=[f_{i1} , f_{i2}]$ is
a row of size 2.
Transmitted signals are thus:
\begin{equation*}
\begin{split}
&x_1(2n) = \textbf{F}_1 \: [s_1(n),s_2(n-1)]^T = f_{11} s_1(n) + f_{12} s_2(n-1)\\
&x_2(2n+1) = \textbf{F}_2 \: [s_1(n),s_2(n)]^T = f_{21} s_1(n) +
f_{22} s_2(n)
\end{split}
\end{equation*}

In Linear NC cooperation scheme, the power constraint
becomes $P_i=\varepsilon_s \|\textbf{F}_i\|_F^2 \leq \frac{2P}{W}$.
We will consider precoding functions such that $\|\textbf{F}_i\|_F^2
= 1$,
i.e. $f_i$ does not increase
the power transmitted by source $S_i$ but shares it between the
source message and the relayed message.

\textbf{Remark :} orthogonal TDMA transmissions without relaying can
be seen as a particular case of network coding where
$\textbf{F}_1=[1,0]$ and $\textbf{F}_2=[0,1]$. Orthogonal
interference-free cooperation \cite{Laneman-2004} is
also a particular case of our scheme where
$\textbf{F}_1=[1,0]$ and $\textbf{F}_2=[1,0]$ during two blocks, and
then $\textbf{F}_2=[0,1]$ and $\textbf{F}_1=[0,1]$ during the next
two blocks.

\subsection{Dirty Paper Precoding}

Since interference resulting from NC approach is known at the
transmitter, more advanced NC functions can include decoding and
re-encoding with DPC of messages intended to
different destinations \cite{Yu-Cioffi-2001}. In Dirty Paper NC for PDF, $S_1$ decodes the message carried by $s_2(n-1)$ and
re-encodes it using an independent Gaussian codebook. More
precisely, in order to use dirty paper coding, $S_1$ first orders
destinations based on channel knowledge. Then $S_1$ picks a codeword
for the first destination, before choosing a codeword for the second
destination, with full non-causal knowledge of the codeword intended
to first destination. Thus the second destination does not see
interference due to the codeword for the first destination, whereas
the first destination will see the signal intended to the second
destination as interference. The signal transmitted by $S_1$ is the
sum of the two codewords, with power sharing across the two
codewords taking into account channel knowledge. $S_2$ will proceed
the same way in the following block. The ordering of destinations
chosen at each source affects performances.
Transmitted signals thus become:
\begin{equation*}
\begin{split}
&x_1(2n) = f_{11} s_1(n) + f_{12} s_2'(n-1)\\
&x_2(2n+1) = f_{21} s_1'(n) + f_{22} s_2(n)
\end{split}
\end{equation*}
where $f_{ij}^2$ stands for the power allocated by source $S_i$ to
the codeword intended to destination $D_j$, and $s_j'$ is the
independent codeword produced by a source acting as relay after
decoding the message carried by $s_j$.

\section{Performance Analysis}\label{sec:PerfEval}

Average rate, per user and network throughputs as well as outage behavior are analyzed in slow fading channels.

\subsection{Orthogonal interference-free RDF and PDF}

For cooperative channels in fig.
\ref{fig:SimultaneousTxRelay} (b), using
RDF the mutual information between input $s_1$ and output
$y_{D_1}$ at $D_1$ is \cite{Laneman-2004}:
\begin{equation}
\begin{split}
I_{RDF}(s_1;y_{D_1}) = & \frac{1}{2} \min \{ \log (1 + \rho |h_{S_2 S_1}|^2), \\
                  & \log \left(1 + \rho |h_{D_1 S_1}|^2 + \rho |h_{D_1 S_2}|^2 \right) \}
\end{split}
\end{equation}
where the input SNR is $\rho = \varepsilon_s/\sigma^2 = 2 P / (W
\sigma^2)$.
Mutual information $I_{RDF}(s_2;y_{D_2})$ between input $s_2$ and output $y_{D_2}$ at $D_2$ is given similarly.
Half the degrees of freedom are allocated for transmission to a destination - each  destination is passive half of the time - therefore the throughput of the first user is $\frac{1}{2}I_{RDF}(s_1;y_{D_1})$ and the total network throughput using RDF is:
\begin{equation}
C_{RDF} = \frac{1}{2} I_{RDF}(s_1;y_{D_1}) + \frac{1}{2} I_{RDF}(s_2;y_{D_2})
\end{equation}

The outage probability is defined as in \cite{Laneman-2004}:
\begin{equation}\label{eq:OutProbaRDF}
P^{out}_{RDF}(\rho,R)=Pr[I_{RDF}<R]
\end{equation}
where $R$ is by definition the ratio between rate $r$ in bits per second and the number of degrees of freedom utilized by each terminal \cite{Laneman-2004} :
\begin{equation}
R=\frac{r}{W/2}   \mbox{ in b/s/Hz}
\end{equation}

Using PDF, mutual information between $s_1$ and $y_{D_1}$ is \cite{Laneman-bookChap-2006}:
\begin{equation}
\begin{split}
  I_{PDF}(s_1; & y_{D_1}) =  \frac{1}{2} \min \{ \log (1 + \rho |h_{S_2 S_1}|^2), \\
 &  \log (1 + \rho |h_{D_1 S_1}|^2) + \log (1 + \rho |h_{D_1 S_2}|^2 ) \}
\end{split}
\end{equation}
Mutual information $I_{PDF}(s_2;y_{D_2})$ at $D_2$
is also given by a similar formula \cite{Laneman-bookChap-2006}.
The total network throughput of PDF is given by:
\begin{equation}
C_{PDF} = \frac{1}{2} I_{PDF}(s_1;y_{D_1}) + \frac{1}{2} I_{PDF}(s_2;y_{D_2})
\end{equation}
and the outage probability is:
\begin{equation}\label{eq:OutProbaPDF}
P^{out}_{PDF}(\rho,R)=Pr[I_{PDF}<R]
\end{equation}

\subsection{Linear NC RDF}

\begin{figure*}[htbp]
  \begin{centering}
  \subfigure[Per user Throughput of RDF and Linear-NC-RDF]{
    \label{fig:AvgCapaRDF-NCRDF}
    \includegraphics[scale=0.63]{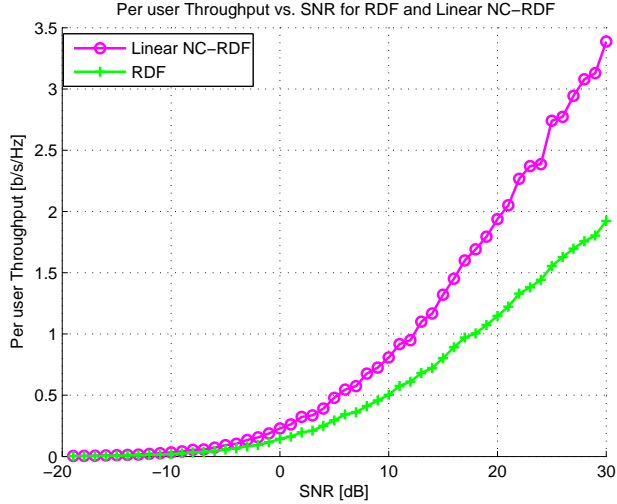}}
   % \hfill
  \subfigure[Per user Throughput of PDF and DPC-NC-PDF]{
    \label{fig:AvgCapaPDF-DPCNCPDF}
    \includegraphics[scale=0.63]{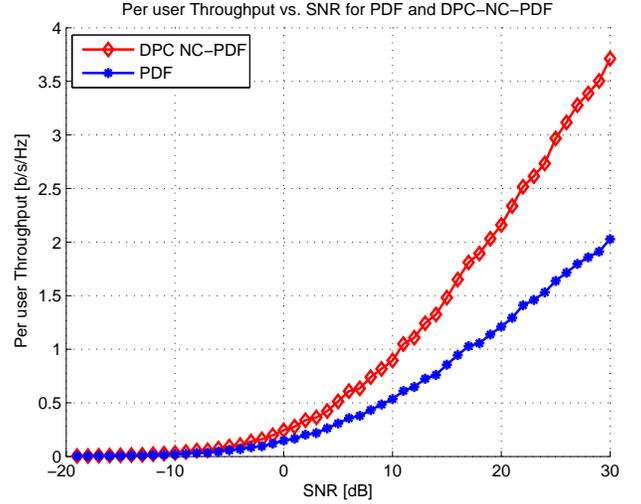}}
    \end{centering}
  \caption{Comparison of Per user Throughputs of classical and NC based cooperative methods}
\end{figure*}

For our proposed network coding cooperative scheme in figure
\ref{fig:SimultaneousTxRelay} (c), when the network coding functions
are linear transformations, mutual information between input $s_1$
and output $y_{D_1}$ at destination $D_1$ can be shown to be:
\begin{equation}\label{eq:NCmutualInfoD1}
\begin{split}
& \!\!\!\! I_{LNC}(s_1;y_{D_1})  = \frac{1}{2} \min  \left\{ \log \left( 1 + \rho |h_{S_2 S_1}f_{11}|^2 \right)  , \right. \\
 %interference free reception at S_2, because it substracts interference
            & \!\!\!\! \left. \log \left(1 + \rho \frac{|h_{D_1 S_1}f_{11}|^2}{1 + \rho |h_{D_1 S_1}f_{12}|^2} + \rho \frac{|h_{D_1 S_2}f_{21}|^2}{1+\rho|h_{D_1 S_2}f_{22}|^2}
                  \right) \right\}
\end{split}
\end{equation}

In the minimum in equation (\ref{eq:NCmutualInfoD1}), the first term
represents the maximum rate at which relay $S_2$ can decode the
source message $s_1$ after canceling the interference known at the
relay (interference is due to the symbol $s_2$ the relay emitted
previously), whereas the second term represents the maximum rate at
which destination $D_1$ can decode given the transmissions from
source $S_1$ and relay $S_2$. A similar formula gives the mutual
information between input $s_2$ and output $y_{D_2}$ at destination
$D_2$, with appropriate changes.
\begin{equation}\label{eq:NCmutualInfoD2}
\begin{split}
& \!\!\!\! I_{LNC}(s_2;y_{D_2})  = \frac{1}{2} \min \left\{ \log \left( 1 + \rho |h_{S_1 S_2}f_{22}|^2 \right)  , \right. \\
                         & \!\!\!\! \left. \log \left(1 + \rho \frac{|h_{D_2 S_2}f_{22}|^2}{1 + \rho |h_{D_2 S_2}f_{21}|^2} + \rho \frac{|h_{D_2 S_1}f_{12}|^2}{1+\rho|h_{D_2 S_1}f_{11}|^2}
                  \right) \right\}
\end{split}
\end{equation}

With Network Coding, all degrees of freedom are used for transmission to each destination. No time is wasted from the destination point of view, thus the throughput for the first user is $I_{LNC}(s_1;y_{D_1})$ and the total network throughput for this strategy is :
\begin{equation}
C_{LNC} = \!\!\!\!\!\! \!\!\!\!\!\! \max_{ \begin{array}{c} \{f_{ij}\}_{ i,j
\in\{1,2\}} \\ |f_{11}|^2+|f_{12}|^2\leq 1 \\
|f_{21}|^2+|f_{22}|^2\leq 1 \end{array} } \!\!\!\!\!\! \!\!\!\!\!\! I
_{LNC}(s_1;y_{D_1}) + I_{LNC}(s_2;y_{D_2})
\end{equation}

The optimization problem turns out to be a non-convex problem, so
that classical convex optimization techniques cannot be used to find
a closed-form expression of the power allocation scheme. Moreover,
because of limitations due to the quality of the link source-relay,
MAC-BC duality \cite{Vishwanath-Jindal-Goldsmith-2003}
cannot be used to solve the
optimization problem as in non-cooperative systems. Finding the
optimal power allocation scheme between transmitted and relayed
signals at each source is different from BC power allocation
problem, because power terms $f_{11}^2$ and $f_{22}^2$ appear in the
capacity of the links between the two sources, first terms in the
minimums in formulas (\ref{eq:NCmutualInfoD1}),
(\ref{eq:NCmutualInfoD2}), (\ref{eq:DPCmutualInfo}), so that the
power allocation scheme maximizing the sum-rates of the two BC
channels between a source and the two destinations may not be the
same as the one maximizing the sum-rate of the cooperative system.

Since all degrees of freedom are utilized by each terminal,
the outage probability is:
\begin{eqnarray}\label{eq:OutProbaLNC}
P^{out}_{LNC}(\rho,R')=Pr[I_{LNC}<R'] \\
\mbox{with } R'=\frac{r}{W} \mbox{ in b/s/Hz}\nonumber
\end{eqnarray}

\subsection{DPC NC PDF}

The mutual information between a source message and the received
signals at the intended destination depends on the two orderings
$\Pi_1, \Pi_2$ of destinations for DPC chosen by both sources.
Since a relay uses an independent codeword to
re-encode the signal it received from the previous source, the
total network throughput for this cooperation scheme belonging to the
family of PDF can be written :
\begin{equation*}
C_{DPC}  = \!\!\!\!\!\! \max_{
\begin{array}{c}\Pi_1,\Pi_2,\{f_{ij}\}_{ i,j \in\{1,2\}}\\
|f_{11}|^2+|f_{12}|^2\leq 1
\\|f_{21}|^2+|f_{22}|^2\leq 1 \end{array}} \!\!\!\!\!\!  I_{DPC}(s_1;y_{D_1}) +
I_{DPC}(s_2;y_{D_2})
\end{equation*}
%with :
\begin{equation}\label{eq:DPCmutualInfo}
\begin{split}
I_{DPC}(s_1; & y_{D_1})  = \frac{1}{2} \min  \left\{ \log \left( 1 + \rho |h_{S_2 S_1}f_{11}|^2 \right)  , \right. \\ %interference free reception at S_2, because it substracts interference
                         & \left. \log (1 + SINR_{11}) + \log (1 + SINR_{21} )
                         \right\} \\
I_{DPC}(s_2; & y_{D_2}) =  \frac{1}{2} \min \left\{ \log \left( 1 + \rho |h_{S_1 S_2}f_{22}|^2 \right)  , \right. \\ %interference free reception at S_2, because it substracts interference
                         & \left. \log (1 + SINR_{12}) + \log (1 + SINR_{22} )
                         \right\}
\end{split}
\end{equation}
where $SINR_{ij}$ is the Signal-to-Interference plus Noise ratio
resulting from the signal transmitted by $S_i$ at
$D_j$:
\begin{equation*}%\label{eq:DPCSINR}
SINR_{ij}  =  \left\{
\begin{array}{c}
    \rho |h_{D_j S_i}f_{ij}|^2 \mbox{ , if } S_i \mbox{ does DPC in favor of } D_{j}\\
    \frac{\rho |h_{D_j S_i}f_{ij}|^2}{1 + \rho |h_{D_j S_i}f_{i\,\bar{j}}|^2} \mbox{ , if } S_i \mbox{ does DPC in favor of } D_{\bar{j}}
\end{array}
             \right.
\end{equation*}

\begin{figure}[htbp]
  \begin{center}
  \includegraphics[scale=0.63]{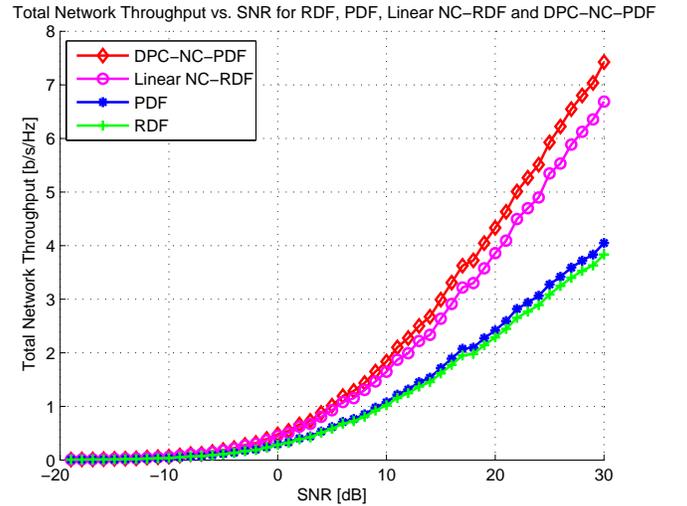}\\
  \caption{Total Network Throughputs of RDF, PDF, linear NC-RDF and DPC-NC-PDF}
  \label{fig:AvgSumRate}
  \end{center}
\end{figure}

The outage probability is defined as
\begin{equation}\label{eq:OutProbaDPC}
P^{out}_{DPC}(\rho,R')=Pr[I_{DPC}<R']
\end{equation}

\begin{figure*}[htbp]
  \begin{centering}
  \subfigure[CDF of Spectral Efficiency - SNR = 10 dB]{
    \label{fig:CDFThroughput-highSNR}
    \includegraphics[scale=0.63]{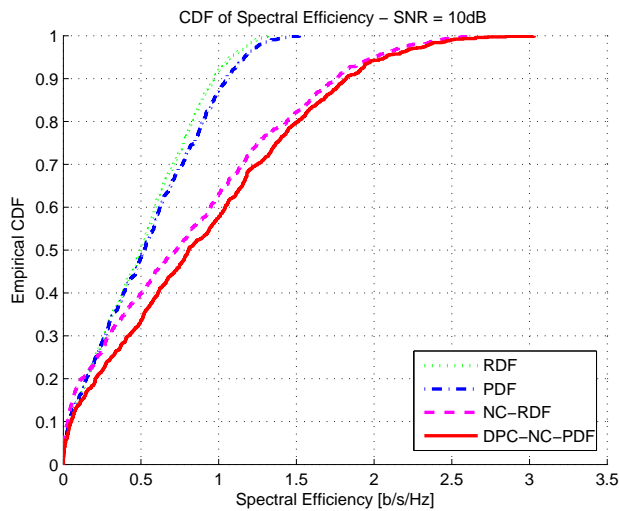}}
     % \hfill
  \subfigure[Outage Probabilities versus SNR]{
    \label{fig:OutageProba}
    \includegraphics[scale=0.63]{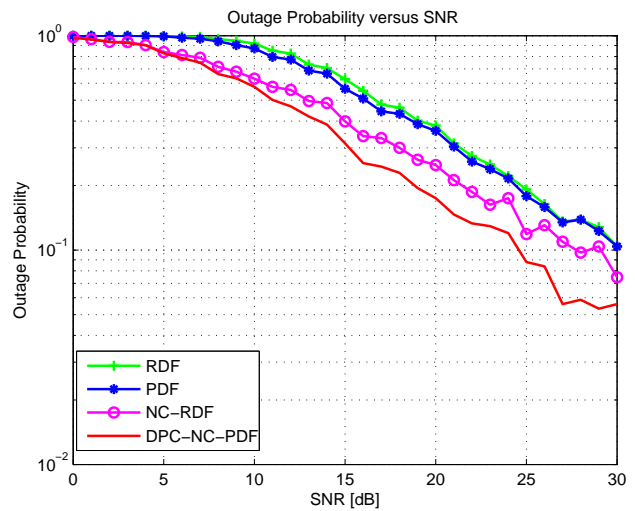}}
    \end{centering}
  \caption{Comparison of CDF of Spectral Efficiencies of classical and NC based cooperative methods}
\end{figure*}

\section{Numerical Results}\label{sec:NumRes}

In this section, numerical results are presented to compare the
different cooperation strategies. Fig.
\ref{fig:AvgCapaRDF-NCRDF}, \ref{fig:AvgCapaPDF-DPCNCPDF} and
(\ref{fig:AvgSumRate}) illustrate average per user throughput and
total network throughput obtained through Monte Carlo Simulations, in the
case of symmetric networks, i.e. in which the fading variances are
identical $\sigma_{v u}^2=1$. Optimal power allocations and orderings $\Pi_i$ were obtained numerically. The average individual throughput are
the same for both users, since they are assumed to have the same
power constraints and the network is symmetric.
Fig. \ref{fig:CDFThroughput-highSNR} and \ref{fig:OutageProba} show the outage behavior of the different strategies.

\subsection{Average Throuhputs}

Fig. \ref{fig:AvgCapaRDF-NCRDF} compares RDF \cite{Laneman-2004}
and LNC for RDF that we propose, and shows that our technique based
on Linear Network coding performs much better thanks to a more
efficient use of spectral resources as well as power resources.
Fig. \ref{fig:AvgCapaPDF-DPCNCPDF} plots the per user throughputs for PDF
\cite{Laneman-2004} and our DPC-NC for PDF.  Once again, the NC
based strategy enhances performances in terms of individual
throughput.

Finally fig. (\ref{fig:AvgSumRate}) allows to compare the total network throughput
of all techniques, and shows the neat improvements in the
network performances thanks to NC methods. Thanks to smart power
sharing between own and relayed signals, even with repetition
coding, and increased spectral efficiency, Linear NC enhances
considerably performances compared to classical RDF and PDF. Using a
more advanced coding technique, DPC to mitigate interferences
generated at destination by the NC methods leads to even better
results.

\subsection{Outage Behavior}

Fig. \ref{fig:CDFThroughput-highSNR} plots the cumulative distribution functions of the per user throughputs. Indeed $$P^{out}_{RDF}(\rho,R)=Pr[I_{RDF}<R]=Pr[I_{RDF}/2<R']$$
Recalling that $I_{RDF}/2$ is the per user throughput, analyzing the outage behavior of the different strategies for a target rate $r$ is equivalent to comparing the CDF of the per user throughputs for a rate value $R'$. A neat improvement in the outage probability is visible in fig.  \ref{fig:CDFThroughput-highSNR} when using network coding cooperation.

Fig. \ref{fig:OutageProba} shows  the outage probabilities (\ref{eq:OutProbaRDF}), (\ref{eq:OutProbaPDF}), (\ref{eq:OutProbaLNC}) and (\ref{eq:OutProbaDPC}), versus the SNR for the different strategies, and a target rate $r=1 b/s $. They illustrate in particular the large energy savings that NC based cooperative strategies allow to reach a target rate.

\section{Conclusion}\label{sec:Conclusion}

Inspired by network coding, we proposed new cooperative strategies
for ad hoc networks,
which improve spectral efficiency of the
cooperative system by relaxing the orthogonality constraint, though
preserving the practical half-duplex constraint. The introduction of
interferences between source and relayed messages, when considering
non-orthogonal transmission scheme, is mitigated thanks to precoding
at transmitter.
We presented two precoding approaches, linear NC
with RDF and Dirty-Paper NC with PDF, relevant technique since the
transmitter knows the interference.
Thanks to precoding, linear or Dirty Paper based, the cost of
the NC approach - introduction of interferences - is less than the
resulting gain in terms of spectral efficiency and performance
analysis shows great improvements in terms of sum-rate capacity over
classical RDF / PDF cooperative strategies.
Future work may include
development of a selective strategy to circumvent limitations due to link
source-relay, extension to multiple-antenna terminals, in particular
assessing how beamforming can improve performances, and last but not least
extension to a large network with several source-destination pairs.

\section*{Acknowledgment}
The authors would like to thank Samsung Advanced Institute of Technology, South Korea and the French Defense Body, DGA for supporting the work of Nadia Fawaz.

% references section

\bibliographystyle{./Biblio/IEEEtran}
\bibliography{./Biblio/IEEEabrv,./Biblio/bibNetworkCoding}

\begin{thebibliography}{10}
\providecommand{\url}[1]{#1}
\csname url@rmstyle\endcsname
\providecommand{\newblock}{\relax}
\providecommand{\bibinfo}[2]{#2}
\providecommand\BIBentrySTDinterwordspacing{\spaceskip=0pt\relax}
\providecommand\BIBentryALTinterwordstretchfactor{4}
\providecommand\BIBentryALTinterwordspacing{\spaceskip=\fontdimen2\font plus
\BIBentryALTinterwordstretchfactor\fontdimen3\font minus
  \fontdimen4\font\relax}
\providecommand\BIBforeignlanguage[2]{{%
\expandafter\ifx\csname l@#1\endcsname\relax
\typeout{** WARNING: IEEEtran.bst: No hyphenation pattern has been}%
\typeout{** loaded for the language `#1'. Using the pattern for}%
\typeout{** the default language instead.}%
\else
\language=\csname l@#1\endcsname
\fi
#2}}

\bibitem{Laneman-2004}
J.~N. Laneman, D.~N.~C. Tse, and G.~W. Wornell, ``Cooperative diversity in
  wireless networks: Efficient protocols and outage behavior,'' \emph{{IEEE}
  Trans. Inform. Theory}, vol.~50, no.~12, pp. 3062--3080, Dec. 2004.

\bibitem{Chou-tutorial}
\BIBentryALTinterwordspacing
P.~Chou. (2006, Mar.) Network coding for the internet and wireless networks.
  Tutorial. University of Michigan. [Online]. Available:
  \url{http://www.eecs.umich.edu/systems/ChouSeminar.ppt}
\BIBentrySTDinterwordspacing

\bibitem{Costa-1983}
M.~Costa, ``Writing on dirty paper (corresp.),'' \emph{{IEEE} Trans. Inform.
  Theory}, vol.~29, no.~3, pp. 439--441, Mar. 1983.

\bibitem{Sendonaris-Erkip-2003}
A.~Sendonaris, E.~Erkip, and B.~Aazhang, ``User cooperation diversity, part i,
  ii,'' \emph{{IEEE} Trans. Commun.}, vol.~51, no.~11, pp. 1927--1948, Nov.
  2003.

\bibitem{Hunter-Sanayei-2006}
T.~E. Hunter, S.~Sanayei, and A.~Nosratinia, ``Outage analysis of coded
  cooperation,'' \emph{{IEEE} Trans. Inform. Theory}, vol.~52, no.~2, pp.
  375--391, Feb. 2006.

\bibitem{Azarian-ElGamal-2005}
K.~Azarian, H.~E. Gamal, and P.~Schniter, ``On the achievable
  diversity-multiplexing tradeoff in half-duplex cooperative channels,''
  \emph{{IEEE} Trans. Inform. Theory}, vol.~51, no.~12, pp. 4152--4172, Dec.
  2005.

\bibitem{Ng-Goldsmith-2004}
C.~T.~K. Ng and A.~J. Goldsmith, ``Transmitter cooperation in ad-hoc wireless
  networks: Does dirty-paper coding beat relaying?'' in \emph{Proc. {IEEE}
  Information Theory Workshop 2004}, Oct. 2004, pp. 277--282.

\bibitem{Lo-2005}
C.~K. Lo, S.~Vishwanath, and R.~W.~J. Heath, ``Rate bounds for {MIMO} relay
  channels using precoding,'' in \emph{Proc. {IEEE} GLOBECOM '05}, vol.~3,
  Nov./Dec. 2005, pp. 277--282.

\bibitem{HostMadsen-2006}
A.~Host-Madsen, ``Capacity bounds for cooperative diversity,'' \emph{{IEEE}
  Trans. Inform. Theory}, vol.~52, no.~4, pp. 1522--1544, Apr. 2006.

\bibitem{Fawaz-Gesbert-Debbah-WSIT2007}
N.~Fawaz, D.~Gesbert, and M.~Debbah, ``When network coding and dirty paper
  coding cooperate,'' in \emph{Proc. {IEEE} Winterschool on Coding and
  Information Theory 2007}, Mar. 2007, p.~63.

\bibitem{Katti-Maric-Goldsmith-2007}
S.~Katti, I.~Mari\'{c}, A.~Goldsmith, D.~Katabi, and M.~M\'{e}dard, ``Joint
  relaying and network coding in wireless networks,'' in \emph{Proc. {IEEE}
  ISIT '07}, June 2007.

\bibitem{Yu-Cioffi-2001}
W.~Yu and J.~M. Cioffi, ``Trellis precoding for the broadcast channel,'' in
  \emph{Proc. {IEEE} GLOBECOM '01}, vol.~2, Nov. 2001, pp. 1344--1348.

\bibitem{Laneman-bookChap-2006}
J.~N. Laneman, \emph{Cooperation in Wireless Networks: Principles and
  Applications}.\hskip 1em plus 0.5em minus 0.4em\relax Springer, 2006, ch.
  Cooperative Diversity: Models, Algorithms, and Architectures.

\bibitem{Vishwanath-Jindal-Goldsmith-2003}
S.~Vishwanath, N.~Jindal, and A.~Goldsmith, ``Duality, achievable rates, and
  sum-rate capacity of gaussian mimo broadcast channels,'' \emph{{IEEE} Trans.
  Inform. Theory}, vol.~49, no.~10, pp. 2658-- 2668, Oct. 2003.

\end{thebibliography}

\end{document}